\documentclass{emulateapj}
\usepackage{color} 

\usepackage{epsfig}
\usepackage{subfigure}
\usepackage{graphicx}
\usepackage{amsmath}
\usepackage{natbib}
\newcommand{\pa}{\partial}
\newcommand{\mb}{\boldsymbol}

\newcommand{\ovl}{\overline}

\shorttitle{Magnetic Flux Concentration in MRI}
\shortauthors{Bai \& Stone}

\begin{document}


\title{Magnetic Flux Concentration and Zonal Flows in Magnetorotational Instability Turbulence}


\author{Xue-Ning Bai\altaffilmark{1,3} \& James M. Stone\altaffilmark{2}}

\altaffiltext{1}{Institute for Theory and Computation,
Harvard-Smithsonian Center for Astrophysics, 60 Garden St., MS-51, Cambridge, MA 02138}
\altaffiltext{2}{Department of Astrophysical Sciences, Peyton Hall, Princeton
University, Princeton, NJ 08544}
\altaffiltext{3}{Hubble Fellow}
\email{xbai@cfa.harvard.edu}




\begin{abstract}
Accretion disks are likely threaded by external vertical magnetic flux, which
enhances the level of turbulence via the magnetorotational instability (MRI). Using
shearing-box simulations, we find that such external magnetic flux also strongly
enhances the amplitude of banded radial density variations known as zonal flows.
Moreover, we report that vertical magnetic flux is strongly concentrated toward
low-density regions of the zonal flow.  Mean vertical magnetic field can be more than
doubled in low-density regions, and reduced to nearly zero in high density regions
in some cases. In ideal MHD, the scale on which magnetic flux concentrates can
reach a few disk scale heights. In the non-ideal MHD regime with strong ambipolar
diffusion, magnetic flux is concentrated into thin axisymmetric shells at some
enhanced level, whose size is typically less than half a scale height. We show that
magnetic flux concentration is closely related to the fact that the magnetic diffusivity
of the MRI turbulence is anisotropic. In addition to a conventional Ohmic-like
turbulent resistivity, we find that there is a correlation between the vertical velocity and
horizontal magnetic field fluctuations that produces a mean electric field that acts to
anti-diffuse the vertical magnetic flux. The anisotropic turbulent diffusivity has
analogies to the Hall effect, and may have important implications for magnetic flux
transport in accretion disks. The physical origin of magnetic flux concentration may
be related to the development of channel flows followed by magnetic reconnection,
which acts to decrease the mass-to-flux ratio in localized regions. The association
of enhanced zonal flows with magnetic flux concentration may lead to global pressure
bumps in protoplanetary disks that helps trap dust particles and facilitates planet
formation.
\end{abstract}



\section{Introduction}\label{sec:intro}

The magnetorotational instability (MRI, \citealp{BH91}) is considered the
most promising mechanism for triggering turbulence and transporting angular momentum
in accretion disks. The properties of the MRI depend on magnetic field geometry. Without
external field, the MRI serves as a dynamo process that keeps dissipating and
re-generating magnetic fields in a self-sustained manner (e.g.,
\citealp{SHGB96,Davis_etal10,Shi_etal10}). On the other hand, MRI turbulence
becomes stronger when the disk is threaded by external (vertical) magnetic flux
\citep{HGB95,BaiStone13a}. Such external magnetic flux may be generically present in
accretion disks, especially in protoplanetary disks (PPDs), from both observational
\citep{Chapman_etal13,Hull_etal14} and theoretical
\citep{BaiStone13b,Bai13,Simon_etal13b} points of view.

Numerical studies of the MRI turbulence have shown that it tends to generate long-lived
large-scale axisymmetric banded density/pressure variations. They are termed zonal
flows, with geostrophic balance between radial pressure gradients and the Coriolis force
\citep{Johansen_etal09}. In PPDs, zonal flows have the attractive potential to concentrate
dust particles into pressure bumps, which may serve as a promising mechanism for
planetesimal formation \citep{Dittrich_etal13}, and also as dust traps to overcome the rapid
radial drift of mm sized grains \citep{Pinilla_etal12}.

Without external magnetic flux, the existence of zonal flows is robust based on local
shearing-box simulations \citep{Simon_etal12a}, although they are not unambiguously
identified in global simulations \citep{Uribe_etal11,Flock_etal12}. In the presence of net
vertical magnetic flux, enhanced zonal flow has been reported from local shearing-box
simulations in the ambipolar diffusion dominated outer regions of PPDs
\citep{SimonArmitage14}. Such enhanced zonal flow is further found to be associated
with the re-distribution of vertical magnetic flux \citep{Bai14}: flux is concentrated into thin
shells in the low-density regions of the zonal flow, while the high-density regions have
almost zero net vertical magnetic flux (see Figure 8 of \citealp{Bai14}).

\begin{figure*}
    \centering
    \includegraphics[width=180mm]{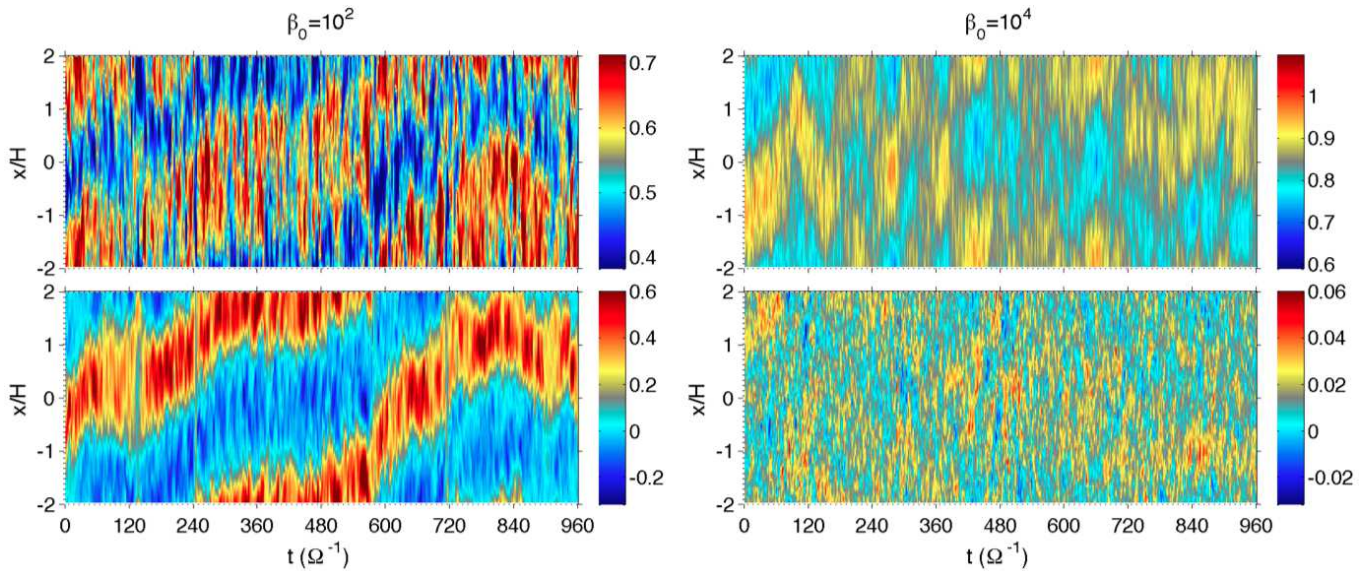}
  \caption{Time evolution of the radial profiles of mean density $\bar{\rho}$ (top) and mean
  vertical magnetic field $\bar{B}_z$ (bottom) in the midplane region from the ideal MHD,
  vertically stratified simulations of \citet{BaiStone13a}. Left and right panels correspond
  to runs B2 (midplane $\beta_0=10^2$) and B4 (midplane $\beta_0=10^4$) in that paper.
  The average is taken azimuthally and vertically within $z=\pm2H$. The color scales are
  centered at the mean value, and span the same range relative to the mean. This Figure
  may be viewed in parallel with the top and bottom panels in Figure 7 of
  \citet{BaiStone13a}.}\label{fig:demo}
\end{figure*}

Magnetic flux concentration by MRI turbulence is evident in earlier shearing-box as well
as global simulations containing net vertical magnetic flux, although it has not been
systematically studied in the literature. For instance, we show in Figure \ref{fig:demo} the
time evolution of radial profiles for the mean gas density $\bar{\rho}$ and the mean vertical
magnetic field $\bar{B}_z$ around disk midplane extracted from runs B2 and B4 in
\citet{BaiStone13a}. These are isothermal ideal magnetohydrodynamic (MHD) stratified
shearing-box simulations of the MRI in the presence of relatively strong net vertical
magnetic flux. The net vertical field is characterized by $\beta_0$, the ratio of gas to
magnetic pressure (of the net vertical field) at the disk midplane, with $\beta_0=100$ and
$10^4$ respectively.
We see from the top panels that strong zonal flows are produced with density variations of
about $30\%$ and $5\%$ around the mean values respectively. The bottom panel shows
the corresponding magnetic flux distribution. Concentration of magnetic flux in low-density
region of the zonal flow is obvious when $\beta_0=10^2$, and the high-density region
contains
essentially zero net vertical magnetic flux. With weaker net vertical field $\beta_0=10^4$,
magnetic flux concentration is still evident but weaker. We emphasize that the systems are
highly turbulent where the level of density and magnetic fluctuations is much stronger than
their mean values (see Figures 3 and 4 of \citealp{BaiStone13a}).

In this work, we systematically explore the phenomenon of magnetic flux concentration by
performing a series of local shearing-box simulations (Section 2), both in the ideal MHD
regime and in the non-ideal MHD regime with ambipolar diffusion as a proxy for the outer
regions of PPDs. All simulations are unstratified and include net vertical magnetic flux. A
phenomenological model is presented in Section 3 to address the simulation results.
Using this model, we systematically explore parameter space in Section 4. While we focus
on unstratified simulations in this work, we have tested that the phenomenological model
can be applied equally well to stratified simulations such as shown in Figure \ref{fig:demo}.
A possible physical mechanism for magnetic flux concentration, together with its astrophysical
implications are discussed in Section 5. We conclude in Section 6.

\section[]{Magnetic Flux Concentration in Shearing-Box Simulations}\label{sec:shear}

We first perform a series of unstratified 3D shearing-box simulations using the Athena
MHD code \citep{Stone_etal08}. The orbital advection scheme \citep{StoneGardiner10}
is always used to remove location-dependent truncation error and increase the time step
\citep{FARGO,Johnson_etal08}.
The MHD equations are written in Cartesian coordinates for a local disk patch in
the corotating frame with angular velocity $\Omega$. With $(x, y, z)$ denoting the radial,
azimuthal and vertical coordinates, the equations read
\begin{equation}
\frac{\pa\rho}{\pa t}+\nabla\cdot(\rho{\mb v})+v_K\frac{\pa\rho}{\pa y}=0\ ,\label{eq:cont}
\end{equation}
\begin{equation}
\frac{\pa\rho{\mb v}}{\pa t}+v_K\frac{\pa\rho{\mb v}}{\pa y}
+\nabla\times(\rho{\mb v}{\mb v}+{\sf T})=
-\frac{1}{2}\rho\Omega v_x{\mb e}_y
+2\rho\Omega v_y{\mb e}_x\ ,\label{eq:momentum}
\end{equation}
\begin{equation}
\frac{\pa{\mb B}}{\pa t}
=-\frac{3}{2}B_x\Omega{\mb e}_y+\nabla\times\bigg[{\mb v}\times{\mb B}
+\frac{({\mb J}\times{\mb B})\times{\mb B}}{\gamma\rho_i\rho}\bigg]
\ ,\label{eq:induction}
\end{equation}
where ${\sf T}\equiv(P+B^2/2){\sf I}-{\mb B}{\mb B}$ is the total stress tensor,
$\rho$, $P$, $v_K$, ${\mb v}$ and ${\mb B}$ denote gas density, pressure,
background Keplerian velocity, background subtracted velocity, and magnetic field,
respectively. We adopt an isothermal equation of state $P=\rho c_s^2$ with $c_s$
being the sound speed. The unit for magnetic field is such that magnetic permeability
$\mu=1$, and ${\mb J}=\nabla\times{\mb B}$ is the current density. The disk scale
height is defined as $H\equiv c_s/\Omega$. We set $\rho_0=\Omega=c_s=H=1$
in code units, where $\rho_0$ is the mean gas density. The last term in the induction
equation is due to ambipolar diffusion (AD), with $\gamma$ being the coefficient for
momentum transfer in ion-neutral collisions, and $\rho_i$ is the ion density. The
strength of AD is measured by the Elsasser number $Am\equiv\gamma\rho_i/\Omega$,
the frequency that a neutral molecule collides with the ions normalized to the disk orbital
frequency \citep{ChiangMurrayClay07}. We consider both the ideal MHD regime, which
corresponds to $Am\rightarrow\infty$, and the non-ideal MHD regime with
$Am\sim1$, appropriate for the outer regions of PPDs \citep{Bai11a,Bai11b}.

\begin{figure*}
    \centering
    \includegraphics[width=180mm]{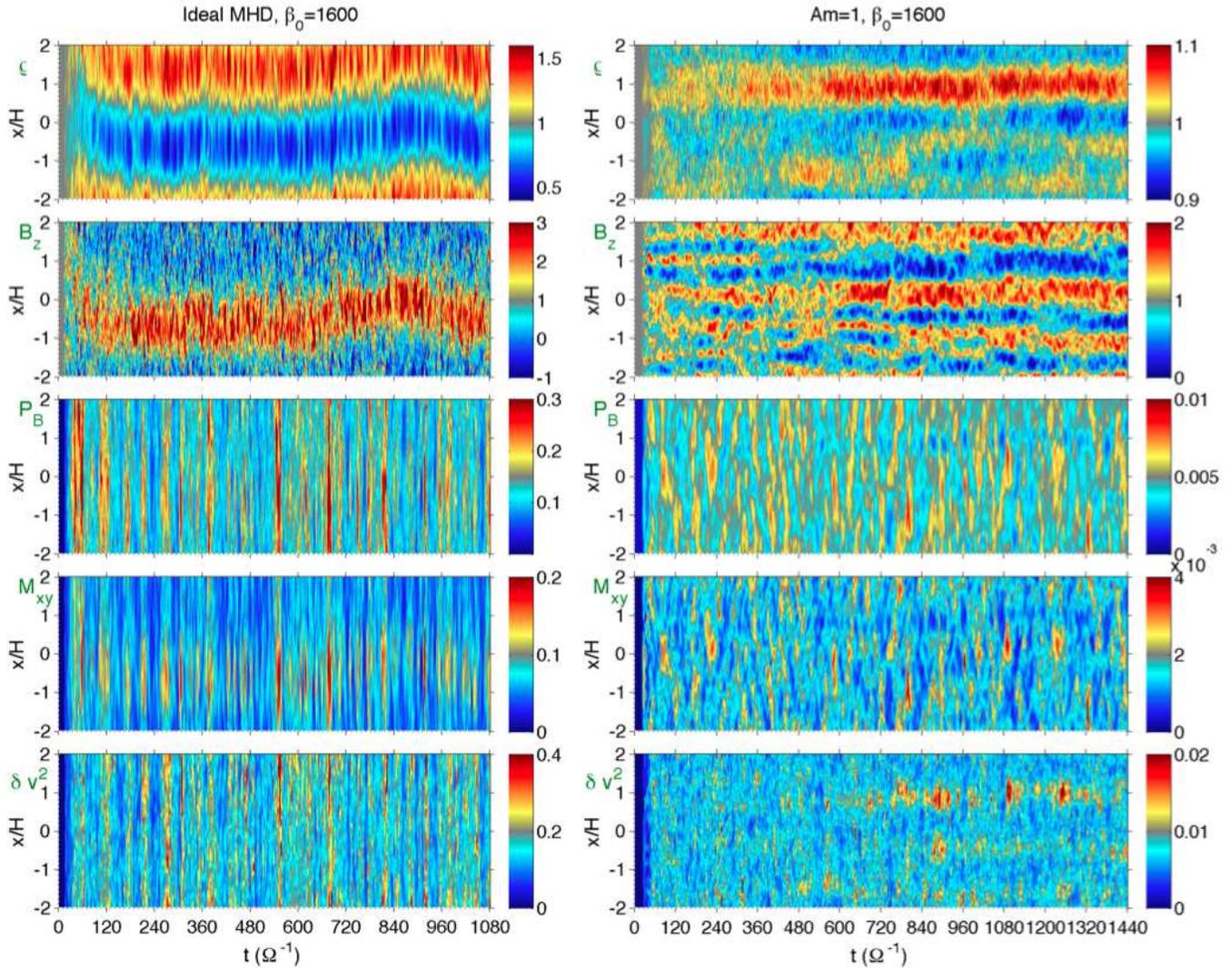}
  \caption{Time evolution of the radial profiles of main diagnostic quantities from our
  fiducial ideal MHD run ID-4-16 (left) and fiducial non-ideal MHD run AD-4-16 (right).
  For each run, from top to bottom, we show the evolution of mean density $\bar{\rho}$,
  normalized mean vertical magnetic field $\bar{B}_z/B_{z0}$, magnetic pressure $P_B$,
  Maxwell stress $M_{xy}$, and turbulent velocity fluctuation
  $\delta v^2=\delta v_x^2+\delta v_z^2$.}\label{fig:shbox-fid}
\end{figure*}

All our simulations contain net vertical magnetic field $B_{z0}$, measured by initial
plasma $\beta_0=2P_0/B_{z0}^2$, the ratio of gas pressure to the magnetic pressure
of the net vertical field. We perform a total of 9 runs listed in Table \ref{tab:shbox}.
Typical simulation run time ranges from $T=1080\Omega^{-1}$ ($\sim172$ orbits) to
$T=2700\Omega^{-1}$ ($\sim430$ orbits).
Physical run parameters include $\beta_0$ and $Am$, while numerical parameters include
simulation box size and resolution. Fiducially, we adopt box size of
$L_x\times L_y\times L_z=4H\times4H\times H$, resolved with
$192\times96\times48$ cells for ideal MHD simulations.
For non-ideal MHD simulations, we increase the resolution to $256\times128\times64$ cells,
which helps better resolve the MRI turbulence (see discussion below). We also explore the
effect of horizontal domain size by varying $L_x$ from $2H$ to $12H$ while keeping
the same resolution (and $L_y=\max[L_x, 4H]$). We set $\beta_0=1600$ as the standard
value, but we also consider $\beta_0=400$ and $6400$ for comparison.

All simulations quickly saturate into the MRI turbulence in a few orbits. Standard
diagnostics of the MRI include the Maxwell stress
\begin{equation}
M_{xy}\equiv-B_xB_y\ ,
\end{equation}
and the Reynolds stress $\rho v_xv_y$.
Their time and volume averaged values normalized by pressure give the
Shakura-Sunyaev parameters $\alpha_{\rm Max}$ and $\alpha_{\rm Rey}$ respectively. 
In Table \ref{tab:shbox}, we list these values for all our simulations, averaged from
$t=360\Omega^{-1}$ onward. Also listed is the plasma $\beta$ parameter, ratio of gas to
magnetic pressure at the saturated state. We see that in both ideal and non-ideal MHD runs,
$\alpha_{\rm Max}$ and $\alpha_{\rm Rey}$ increases with net vertical magnetic flux, as is
well known \citep{HGB95,BaiStone11}. Also, they all roughly satisfy the empirical relation
$\alpha\beta\approx1/2$ in both ideal MHD and non-ideal MHD cases
\citep{Blackman_etal08,BaiStone11}, where $\alpha=\alpha_{\rm Max}+\alpha_{\rm Rey}$.

To ensure that our simulations have sufficient numerical resolution, we have computed the
quality factor $Q_z\equiv\lambda_{\rm MRI}/\Delta z$ \citep{Noble_etal10}, where
$\lambda_{\rm MRI}$ is the characteristic MRI wavelength based on the total (rms) vertical
magnetic field strength. For the most unstable wavelength in ideal MHD, we have
$\lambda_{\rm MRI}=9.18\beta_z^{-1/2}$ \citep{HGB95}, where $\beta_z=2P/\ovl{B_z^2}$ is
the plasma $\beta$ parameter for the vertical field component. In non-ideal MHD with
$Am=1$, we find $\lambda_{\rm MRI}=17.47\beta_z^{-1/2}$ \citep{BaiStone11}. Similarly,
one can define $Q_y\equiv\lambda_c/\Delta y$, where $\lambda_c$ is defined the same
way as $\lambda_{\rm MRI}$ but using $\beta_\phi$ instead of $\beta_z$. In general, the MRI
is well resolved when $Q_y\gtrsim20$ and $Q_z\gtrsim10$ \citep{Hawley_etal11}. We find
that in all our simulations are well resolved based on this criterion. Further details are provided
in Section \ref{sec:param}.

In Figure \ref{fig:shbox-fid}, we show the time evolution of the radial profiles of various diagnostic
quantities for our fiducial ideal and non-ideal MHD runs. The results are discussed below.
Other runs will be discussed in Section \ref{sec:param}.

\subsection[]{The Ideal MHD Case}\label{ssec:fid1}

In this ideal MHD run, we see that a very strong zonal flow is produced, with density contrast
up to $50\%$. In the mean time, there is a strong anti-correlation between gas density and
mean vertical magnetic field, with most magnetic flux concentrated in the low density regions.
In this fiducial run with radial box size $L_x=4H$, there is just one single ``wavelength" of
density and mean field variations. The phase of the pressure maxima drifts slowly in a
random way over long timescales, accompanied by a slow radial drift of the mean field profile;
but overall, the system achieves a quasi-steady-state in terms of density and magnetic flux
distributions.

Combined with Figure \ref{fig:demo}, we see that both unstratified and stratified shearing-box
simulations show similar phenomenon of magnetic flux concentration and zonal flows. This
fact indicates that the same physics is operating, independent of buoyancy. We stress that the
mean vertical field, even in the highly concentrated region, is much weaker than the rms vertical
field from the MRI turbulence. Therefore, the physics of magnetic flux concentration lies in the
intrinsic properties of the MRI turbulence.

From the last three panels on the left of Figure \ref{fig:shbox-fid}, we see that the action of
the Maxwell stress (which is the driving force of the zonal flow) is bursty. Such behavior
corresponds to the recurrence of the channel flows followed by dissipation due to magnetic
reconnection \citep{SanoInutsuka01}.
The Maxwell stress is most strongly exerted in regions where magnetic flux is concentrated.
Interestingly, magnetic pressure shows bursty behavior similar to the Maxwell stress, but
its strength does not show obvious signs of radial variation. On the other hand, turbulent velocity
in the $x-z$ plane, given by $\delta v^2=\delta v_x^2+\delta v_z^2$, is strongest in regions
with weaker magnetic flux during each burst\footnote{By contrast, we find $\delta v_y^2$ (after
removing the zonal flow) peaks in regions with magnetic fluctuation.}. While we are mostly
dealing with time-averaged quantities in this work, one should keep in mind about such
variabilities on timescales of a few orbits.


\begin{table*}
\caption{List of all shearing-box simulations}\label{tab:shbox}
\begin{center}
\begin{tabular}{c|ccc|ccc|cc|cccc|c}\hline\hline
 Run & Box size ($H$) & $\beta_0$ & $Am$ &  $\alpha_{\rm Max}$ &  $\alpha_{\rm Rey}$
 & $\langle\beta\rangle$ & $\Delta\rho/\rho_0$ & $\ovl B_{z}^{\rm Max}/B_{z0}$ & $\alpha_m$ & $\alpha_t$
 & $Q'$ & $\alpha_{xy}$ & $t\ (\Omega^{-1})$\\\hline

ID-4-4 & $4\times4\times1$ & $400$ & $\infty$ & $0.15$ & $0.055$& $3.6$ & $0.26$ & $1.8$ & $0.22$ & $0.054$ & $-0.11$ & $-0.12$ & $360-660$\\

ID-4-16 & $4\times4\times1$ & $1600$ & $\infty$ & $0.070$ & $0.026$& $7.4$ & $0.37$ & $2.2$ & $0.086$ & $0.033$ & $-0.081$ & $-0.072$ & $360-720$\\

ID-4-64 & $4\times4\times1$ & $6400$ & $\infty$ & $0.034$ & $0.010$& $14$ & $0.23$ & $2.1$ & $0.032$ & $0.010$ & $-0.050$ & $-0.033$  & $600-780$\\\hline

ID-2-16 & $2\times4\times1$ & $1600$ & $\infty$ & $0.083$ & $0.022$& $5.7$ & $0.014$ & $1.33$ & $--$ & $--$ & $---$ & $--$ & $480-600$\\


ID-8-16 & $8\times8\times1$ & $1600$ & $\infty$ & $0.069$ & $0.025$& $7.1$ & $0.44$ & $2.0$ & $0.014$ & $0.032$ & $-0.089$ & $-0.069$ & $1260-1500$\\


ID-16-16 & $16\times16\times1$ & $1600$ & $\infty$ & $0.070$ & $0.024$& $6.9$ & $0.29$ & $1.4$ & $--$ & $--$ & $--$ & $--$ & $1170-1350$\\\hline

AD-4-16 & $4\times4\times1$ & $1600$ & $1$ & $1.5$E$-3$ & $9.1$E$-4$ & $2.1$E+$2$ & $0.061$ & $1.6$ & $5.6$E$-3$ & $1.1$E$-3$ & $-0.093$ & $2.6$E$-3$ & $1080-1440$\\

AD-4-64 & $4\times4\times1$ & $6400$ & $1$ & $4.8$E$-4$ & $6.0$E$-4$& $5.0$E+$2$ & $0.081$ & $2.6$ & $1.6$E$-3$ & $4.9$E$-4$ & $-0.064$ & $1.2$E$-3$ & $1200-1440$\\

AD-2-64 & $2\times4\times1$ & $6400$ & $1$ & $4.4$E$-4$ & $4.1$E$-4$ & $5.3$E$+2$ & $0.024$ & $2.0$ & $7.6$E$-3$ & $5.8$E$-4$ & $-0.048$ & $1.1$E$-3$ & $210-450$\\

\hline\hline
\end{tabular}
\end{center}

\end{table*}

\subsection[]{The Non-ideal MHD Case}\label{ssec:niruns}

On the right of Figure \ref{fig:shbox-fid}, we show the time evolution of the radial profiles
of main diagnostic quantities from our fiducial non-ideal MHD simulation AD-4-16.
Zonal flows and magnetic flux concentration are obvious from the plots.
One important difference from the ideal MHD case is that the scale that magnetic flux
concentrates is much smaller: we observe multiple shells of concentrated magnetic flux
whose width is around $0.5H$ or less. The shells may persist, split, or merge during the
evolution, while their locations are well correlated with the troughs in the radial density
profile. Many other aspects of the evolution are similar to the ideal MHD case, such as
the action of Maxwell stress, and the distribution of $P_B$ and $\delta v^2$. These
flux-concentrated shells closely resemble the shells of magnetic flux observed in stratified
simulations shown in Figure 8 of \citet{Bai14}. Again, the similarities indicate that the
physics of magnetic flux concentration is well captured in unstratified simulations.


In our unstratified simulations, the zonal flow is weaker than in the ideal MHD cases, where
the amplitude of density variations is typically $10\%$ or less. The level of radial density
variations in stratified simulations is typically larger \citep{SimonArmitage14,Bai14}.
Meanwhile, it appears that magnetic flux concentration is more complete in stratified
simulations: most of the magnetic flux is concentrated into the shells, while other regions
have nearly zero net vertical flux (again see Figure 8 of \citealp{Bai14}). Also, the
flux-concentrated shells are more widely separated in stratified simulations. Note that
these stratified simulations contain an ideal-MHD, more strongly magnetized and fully MRI
turbulent surface layer, which may affect the strength of the zonal flow at disk midplane (via
the Taylor-Proudman theorem) as well as the level of magnetic flux concentration.
Nonetheless, addressing these differences is beyond the scope of this work. 

Finally, we note that the zonal flow and magnetic flux concentration phenomena were already
present in our earlier AD simulations \citep{BaiStone11,Zhu_etal14b}. These simulations
either focused on the Shakura-Sunyaev $\alpha$ parameter, or the properties of the MRI
turbulence, while the radial distribution of magnetic flux was not addressed.


\section[]{A Phenomenological Model}\label{sec:model}

\begin{figure*}
    \centering
    \includegraphics[width=160mm]{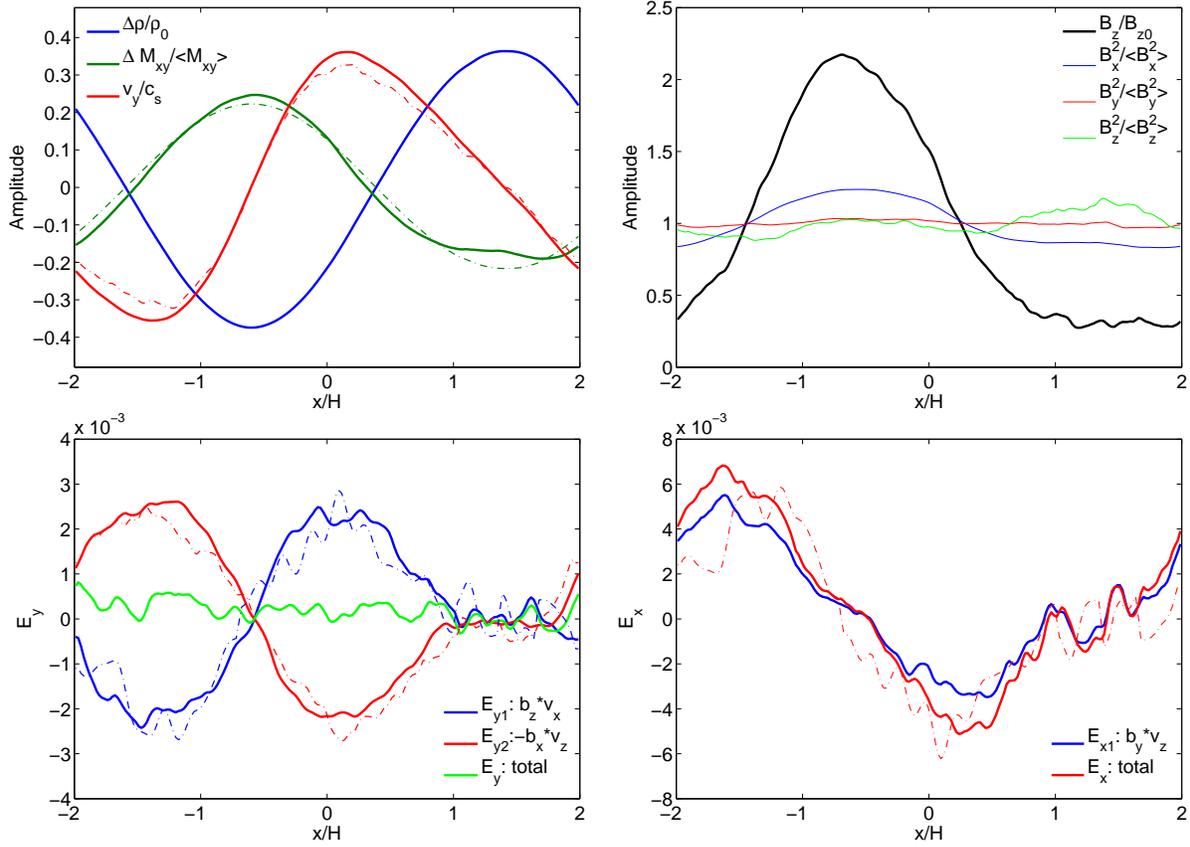}
  \caption{Radial profiles of various quantities in the saturated state of run ID-4-16, as indicated
  in the legends in each panel. Dashed-dotted lines are fits to the measured profiles based on the
  phenomenological model in Section 3.}\label{fig:prof}
\end{figure*}

In this section, we consider our fiducial run ID-4-16 for a detailed case study.
We take advantage of the fact that the system achieves a quasi-steady-state in its radial
profiles of density and magnetic flux, and construct a phenomenological, mean-field
interpretation on magnetic flux concentration and enhanced zonal flows.
We use overbar to denote quantities averaged over the $y-z$ dimensions (and certain period of
time), which have radial dependence. We use $\langle\cdot\rangle$ to represent time and volume
averaged values in the entire simulation domain at the saturated state of the MRI turbulence.
In Figure \ref{fig:prof} we show the radial profiles of some main diagnostic quantities. They are
obtained by averaging from $t=360\Omega^{-1}$ to $720\Omega^{-1}$, where the density and
magnetic flux profiles approximately maintain constant phase. Detailed analysis are
described below.

\subsection[]{Force balance}

Zonal flow is a result of geostrophic balance between radial pressure gradient and the Coriolis
force
\begin{equation}
\frac{c_s^2}{2\Omega}\frac{\pa\ovl \rho}{\pa x}=\ovl\rho\ovl v_y\approx\rho_0\ovl v_y\ .
\end{equation}
From the top left panel of Figure \ref{fig:prof}, we see that the above formula accurately fits
the measured profile of $\ovl v_y$.

The pressure gradients are driven by radial variations of the Maxwell stress, balanced by mass
diffusion
\citep{Johansen_etal09}. Using $D_m$ to denote the mass diffusion coefficient, one obtains
\begin{equation}
\frac{2}{\Omega}\frac{\pa\ovl M_{xy}}{\pa x}=-D_m\frac{\pa\ovl\rho}{\pa x}\ .\label{eq:massdiff}
\end{equation}
Therefore, in a periodic box, the density variation should be anti-correlated with the Maxwell stress.
Asserting $D_m\equiv\alpha_mc_sH$ and assuming $\alpha_m$ is a constant, we obtain
\begin{equation}\label{eq:alpham}
\frac{\Delta\ovl M_{xy}}{\langle M_{xy}\rangle}\approx-\frac{\alpha_m}{2\alpha_{\rm Max}}\frac{\Delta\ovl\rho}{\rho_0}\ ,
\end{equation}
where $\Delta\ovl A\equiv\ovl{A}-\langle A\rangle$ for any quantity $A$.

We can fit the mass diffusion coefficient based on Equation (\ref{eq:alpham}), and obtain
$\alpha_m\approx1.2\alpha_{\rm Max}\approx0.086$. Also from the top left panel of
Figure \ref{fig:prof}, we see that the fitting result agrees extremely well with the measured
profile of $\ovl M_{xy}$.

\subsection[]{Magnetic Flux Evolution}\label{ssec:evolveB}

The evolution of vertical magnetic flux is controlled by the toroidal electric field via the
induction equation (\ref{eq:induction})
\begin{equation}
\frac{\pa\ovl{B}_z(x)}{\pa t}=-\frac{\pa\ovl{E}_y}{\pa x}\ ,
\end{equation}
where in ideal MHD, the toroidal electric field can be decomposed into
\begin{equation}
\overline{E}_y=\ovl{E}_{y1}+\ovl{E}_{y2}=\ovl{v_xB_z}-\ovl{v_zB_x}\ .\label{eq:Ey}
\end{equation}
In the above, the first term describes the advective transport of magnetic flux
by turbulent resistivity
\begin{equation}
\overline{E}_{y1}=\ovl{v_xB_z}\approx\eta_t\ovl{J}_y=-\eta_t\pa_x\ovl{B}_z\ ,\label{eq:eta_t}
\end{equation}
where $\eta_t\equiv\alpha_tc_sH$ is the turbulent resistivity. The outcome is that
accumulation of magnetic flux tends to be smeared out. We can fit the value of
$\alpha_t$ from the profiles of $\ovl B_z$ and $\ovl E_{y1}$ to obtain
$\alpha_t\approx0.033$, which is the same order as $\alpha_{\rm Max}$. While
the data are somewhat noisy, we see from the bottom left panel of Figure \ref{fig:prof}
that the profile of $\ovl E_{y1}$ is well fitted from Equation (\ref{eq:eta_t}). This is the
basic principle for measuring turbulent resistivity from the MRI
\citep{GuanGammie09,LesurLongaretti09,FromangStone09}.

The second term in (\ref{eq:Ey}) describes the generation of vertical field by tilting the
radial field. Since we expect ${\ovl v}_z=0$ and ${\ovl B}_x=0$
in the MRI turbulence, its contribution must come from a correlation between $v_z$
and $B_x$, which is primarily responsible for magnetic flux accumulation.
The fact that the system achieves a quasi-equilibrium state indicates that their sum
$\ovl E_y\approx0$. Therefore, contribution from $\ovl E_{y2}$ must balance the
turbulent diffusion term $\ovl E_{y1}$. This is indeed the case, as we see from Figure
\ref{fig:prof}.

\subsection[]{Turbulent Diffusivity}\label{ssec:diff}

The saturated state of the system has a mean toroidal current $\ovl J_y=-\pa\ovl B_z/\pa x$
but zero mean toroidal electric field $\ovl E_y\approx0$. Applying an isotropic Ohm's law to
the system would yield infinite conductivity. This is obviously not the case. The issue can be
resolved if the turbulent conductivity/diffusivity is {\it anisotropic} with strong off-diagonal
components.

More generally, we write
\begin{equation}
\ovl E_i=\eta_{ik}\ovl J_k\ ,\label{eq:aniso}
\end{equation}
where $i, j, k$ denote any of the $x, y, z$ components, and one sums over index $k$.
Given the mean $\ovl J_y$, we have analyzed all other components of the mean electric
field. We find that the mean vertical electric field $\ovl E_z$ is consistent with zero, while
there is a non-zero mean radial electric field
\begin{equation}
\ovl E_x=\ovl E_{x1}+\ovl E_{x2}=\ovl{v_zB_y}-\ovl{\delta v_y\delta B_z}\ .
\end{equation}
Note that in the second term $\ovl E_{x2}$, we have removed the component $\ovl v_y\ovl B_z$,
which corresponds to the advection of vertical field due to disk rotation and is physically
irrelevant to the MRI turbulence.

In the bottom right panel of Figure \ref{fig:prof}, we show the radial profiles of $\ovl E_x$ and
$\ovl E_{x1}$.  We see that $\ovl E_x$ is approximately in phase with $-\ovl E_{y1}$ and
$\ovl E_{y2}$. This observation indicates that at the saturated state, the system is
characterized by an anisotropic turbulent diffusivity that is off-diagonal, given by
\begin{equation}
\ovl E_x\approx\eta_{xy}\ovl J_y\ .
\end{equation}

We can fit the value of $\eta_{xy}\equiv\alpha_{xy}c_sH$ to obtain
$\alpha_{xy}\approx-0.072\approx-\alpha_{\rm Max}$. In the bottom right panel of Figure
\ref{fig:prof}, we see that although the fitting result is not perfect, it captures the basic
trend on the radial variations of $\ovl E_x$. It is satisfactory since some features can be
smoothed out over the time average due to the (small) phase shift of the density/magnetic
flux profiles.

Anisotropic diffusivities in MRI turbulence have been noted in \citet{LesurLongaretti09},
who measured most components of the diffusivity tensor by imposing some fixed amplitude
mean field variations in Fourier space. In particular, they found that the value of
$\eta_{xy}$ is typically negative\footnote{Note that they used a different coordinate system
from ours. Our $\eta_{xy}$ corresponds to their $-\eta_{yx}$, and our $\eta_t$ corresponds
to their $\eta_{xx}$.}, and the value of $|\alpha_{xy}|$ can be a substantial fraction of
$\alpha_{\rm Max}$. Also, they found that $|\eta_{xy}|$ is typically a factor of several larger
than the diagonal component, which is our equivalence of $\eta_t$. Our measurements of
$\eta_{xy}$ are consistent with their results.

\subsection[]{Connection between Anisotropic Diffusivity and Magnetic Flux Concentration}

We see from the bottom right panel of Figure \ref{fig:prof} that contributions to $\ovl E_x$ is
completely dominated by $\ovl E_{x1}=\ovl{v_zB_y}$, indicating a correlation between $v_z$
and $B_y$. In Section \ref{ssec:evolveB}, we see that magnetic flux concentration is mainly
maintained by $\ovl E_{y2}=-\ovl{v_zB_x}$, indicating an anti-correlation between $v_z$ and
$B_x$. Since $B_y$ and $-B_x$ are correlated in the MRI turbulence (to give the Maxwell
stress $-\ovl{B_xB_y}>0$), it is not too surprising that the two correlations are related to
one another: $\ovl E_{y2}$ and $\ovl E_{x1}$ are in phase as we see in Figure \ref{fig:prof}.

Our analysis suggests that magnetic flux concentration is a direct consequence of the anisotropic
diffusivity/conductivity in the MRI turbulence. In addition to turbulent resistivity given by
$\ovl E_{y1}$, another anisotropic component, resulting from correlations between
$v_z$ and the horizontal magnetic field, contributes to both $\ovl E_{y2}$ and $\ovl E_{x1}$.
The latter exhibits as $\eta_{xy}$, while the former acts to concentrate vertical magnetic flux.

\subsection[]{Analogies to the Hall Effect}

We note that the generation of $\ovl E_x$ from $\ovl J_y$ in the presence of mean vertical
field is analogous to the {\it classical} Hall effect. If we empirically set $\eta_{xy}\equiv QB_{z0}$,
the electric field in the saturated state may be written as
\begin{equation}
\ovl{\mb E}\approx Q\ovl{\mb J}\times\ovl{\mb B}\ .\label{eq:Ehc}
\end{equation}
Since only $\ovl J_y$ and $\ovl B_z$ are non-zero, this leads to a net $\ovl E_x$, consistent
with our measurement.

The analogy above prompts us to draw another analogy between the {\it microscopic} Hall
effect and magnetic flux concentration, which was demonstrated in \citet{KunzLesur13}.
For the microscopic Hall effect, the Hall electric field can be written as
\begin{equation}
\ovl{\mb E}^{h}\approx Q'\ovl{{\mb J}\times{\mb B}}\ .\label{eq:Ehm}
\end{equation}
It generates a mean toroidal electric field via
\begin{equation}\label{eq:Ehy}
\ovl E_y^{h}\approx Q'\ovl{B_x(\pa_x{B_y})}\approx-Q'\frac{\pa\ovl M_{xy}}{\pa x}\ .
\end{equation}
We can fit $\ovl E_{y2}$ using the above relation to obtain $Q'\approx-0.080$ in code unit.
As seen in Figure \ref{fig:prof}, the radial profile of $\ovl E_{y2}$ is fitted very well.
Also note that both $Q$ and $Q'$ are negative based on our fitting results to $\ovl E_x$
and $\ovl E_{y2}$. While different phenomenological considerations are used to arrive
at Equations (\ref{eq:Ehc}) and (\ref{eq:Ehm}), the directionality of the two electric field
components is in line with the Hall-like interpretation.

\begin{figure*}
    \centering
    \includegraphics[width=180mm]{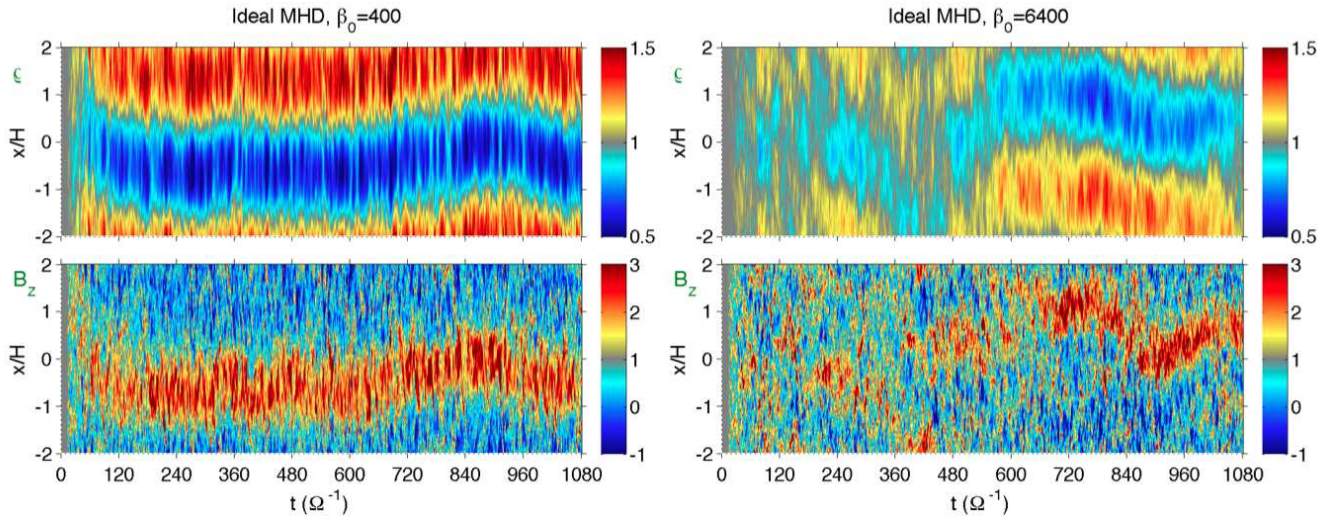}
  \caption{Time evolution of the radial profiles of $\ovl\rho$ (top) and $\ovl B_z/B_{z0}$
  (bottom) from our ideal MHD runs with different $\beta_0$: ID-4-4 (left) and ID-4-64
  (right).}\label{fig:IDhist}
\end{figure*}

\begin{figure*}
    \centering
    \includegraphics[width=180mm]{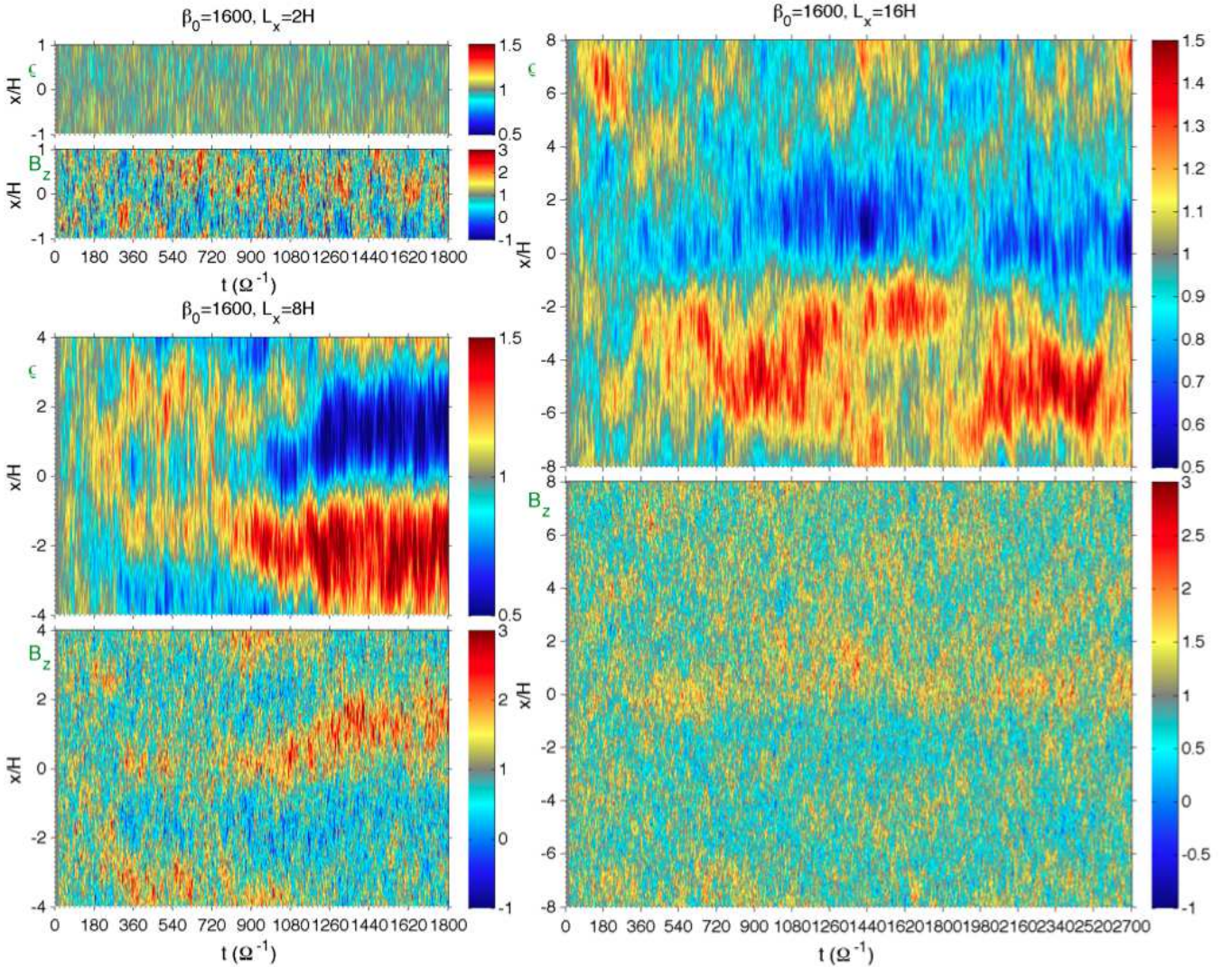}
  \caption{Time evolution of the radial profiles of $\ovl\rho$ and $\ovl B_z/B_{z0}$
  from our ideal MHD runs with different $L_x$: ID-2-16 (top left), ID-8-16 (bottom
  left), ID-16-16 (right).}\label{fig:ID_Lx_hist}
\end{figure*}

Encouraged by the analogies above, if we insert Equation (\ref{eq:Ehy}) as an ansatz for
$\ovl E_{y2}$, then the evolution of vertical magnetic flux can be written as
\begin{equation}\label{eq:Bcon}
\frac{\pa\overline{B}_z}{\pa t}\approx\bigg(\eta_t
+Q'\frac{d\ovl M_{xy}}{d\overline{B}_z}\bigg)\frac{\pa^2\overline{B}_z}{\pa x^2}\ .
\end{equation}
In general, we expect $\ovl M_{xy}$ to increase with net vertical flux until the net
vertical field becomes too strong: $d\ovl M_{xy}/d\ovl B_z>0$
($B_z>0$). From our measurement, we have $Q'<0$. Therefore, the second term
in Equation (\ref{eq:Bcon}) acts as {\it anti}-diffusion of vertical magnetic flux.
It is likely that this term dominates over turbulent diffusion at the initial evolutionary
stage to trigger magnetic flux concentration, while turbulent diffusion catches up
at later stages when sufficient magnetic flux concentration is achieved. Together
with Equation (\ref{eq:alpham}), we see that the gradient of Maxwell stress is
responsible for both launching of the zonal flow and magnetic flux concentration.
This provides a phenomenological interpretation why magnetic flux always
concentrates toward low-density regions. Given our crude phenomenological
treatment, however, we can not provide more detailed descriptions on the flux
concentration process and phase evolution, nor can we explain its saturation
scale and amplitude without involving many unjustified speculations.

Our Equation (\ref{eq:Bcon}) closely resembles Equation (26) of \citet{KunzLesur13},
which were used to explain magnetic flux concentration due to the {\it microscopic}
Hall effect. The counterpart of our $Q'$ in their paper is positive. Therefore, strong
concentration of magnetic flux occurs only when the Hall effect and net vertical
field become sufficiently strong (so that $M_{xy}$ decreases with $B_z$). In our
case, since $Q'$ is negative, magnetic flux concentration is expected even for
relatively weak net vertical field.

\subsection[]{Summary}

In sum, we have decomposed the turbulent diffusivity from the MRI turbulence into
two ingredients. There is a conventional, Ohmic-like turbulent resistivity
$\eta_t$. In addition, we find correlations of $v_z$ with $B_y$ and $B_x$ in the
presence of vertical magnetic flux gradient. The former leads to an anisotropic
diffusivity, which is analogous to the the classical Hall effect. The latter effectively
leads to anti-diffusion of vertical magnetic flux, which is responsible for magnetic flux
concentration, and is analogous to the the microscopic Hall effect.

We emphasize that anisotropic turbulent conductivity/diffusivity is an intrinsic property
of the MRI turbulence. While we draw analogies with the Hall effect, it represents an
phenomenological approach and simply reflects our ignorance about the MRI turbulence.
The readers should not confuse this analogy with the physical (classical or microscopic)
Hall effect, which would lead to polarity dependence (on the sign of $B_{z0}$). Magnetic
flux concentration, on the other hand, has NO polarity dependence.




\section[]{Parameter Exploration}\label{sec:param}

The main results of a series of simulations we have performed to explore parameter space
are summarized in Table \ref{tab:shbox}.
We follow the procedure in Section \ref{sec:model} to analyze the properties related to zonal
flows and magnetic flux concentration. In doing so, we choose a specific time period in each
run where the density and magnetic flux profiles maintain approximately constant phase. They
are listed in the last column of the Table. In many cases, multiple periods can be chosen, and
we confirm that the fitting results are insensitive to period selection.
To characterize the strength of the zonal flow and magnetic flux concentration, we further
include in the table $\Delta\ovl\rho/\rho_0$, the relative amplitude of radial density variations,
and $\ovl B_z^{\rm Max}/B_{z0}$, the ratio of maximum vertical field in the time-averaged
radial profile to its initial background value.
Runs ID-2-16 and ID-16-16 never achieve a quasi-steady state in their density and magnetic
flux distribution, we thus simply measure $\Delta\ovl\rho/\rho_0$ and $\ovl B_z^{\rm Max}/B_{z0}$
over some brief periods, leaving other fitting parameters blank in the Table.

\subsection[]{Ideal MHD Simulations}

All our ideal-MHD simulations have achieved numerical convergence based on the quality
factor criterion discussed in Section \ref{sec:shear}. In particular, in the run with weakest
net vertical field ID-4-64, we find $\ovl Q_y>45$ and $\ovl Q_z>25$ for any $x$, meaning
that the resolution is about twice more than needed to properly resolve the MRI. Runs with
stronger net vertical field give further larger quality factors. Below we discuss the main
simulation results.

\subsubsection[]{Dependence on Net Vertical Field Strength}

We first fix the simulation domain size ($L_x=4H$) and vary the strength of the net vertical
magnetic field to $\beta_0=400$ and $\beta_0=6400$. The time evolution of $\ovl\rho$ and
$\ovl B_z/B_{z0}$ in the two runs are shown in Figure \ref{fig:IDhist}. We see that in general,
enhanced zonal flow requires relatively strong net vertical magnetic field. Our run ID-4-64 with
$\beta_0=6400$ has notably weaker density contrast of $\sim20\%$ compared with
$\sim40\%$ in our fiducial run ID-4-16. It also takes longer time for strong concentration of
magnetic flux to develop. This is in line with the vertically stratified simulations shown in Figure
\ref{fig:demo}. In the limit of zero net vertical flux, the density contrast is further reduced to
$\sim10\%$ \citep{Johansen_etal09,Simon_etal12a}.

For the selected time periods, we find that the phenomenological description in Section
\ref{sec:model} works well of all ideal MHD simulations. There is a systematic trend that the
mass diffusion coefficient $\alpha_m$, turbulent resistivity $\alpha_t$, $Q'$ and $|\alpha_{xy}|$
all increase with increasing net vertical field. In particular, $\alpha_t$ and $\alpha_{xy}$ roughly
scale in proportion with $\alpha_{\rm Max}$.

We do not extend our simulations to further weaker net vertical field, where the MRI would be
under-resolved.
On the other hand, we note that without net vertical magnetic flux, oppositely directed mean
vertical magnetic fields tend to decay/reconnect, rather than grow spontaneously
\citep{GuanGammie09}. Therefore, concentration of vertical magnetic flux occurs only when
there is a net vertical magnetic field threading the disk.

Combining both our unstratified simulation results and the results from stratified simulations of
shown in Figure \ref{fig:demo}, we expect strong concentration of magnetic flux
and enhanced zonal flow to take place for net vertical field $\beta_0\lesssim10^4$.

\subsubsection[]{Dependence on Radial Domain Size}


In our fiducial run ID-4-16, only one single ``wavelength" of density and magnetic flux
variations fit into our simulation box. We thus proceed to perform additional simulations
varying the radial domain size, and show the time evolution of their density and magnetic
flux profiles in Figure \ref{fig:ID_Lx_hist}.

We first notice that when using a smaller box with $L_x=2H$, the zonal structures
become much weaker. They appear to be more intermittent, have finite lifetime, and
undergo rapid and random radial drift. One can still see that magnetic flux is concentrated
toward low density regions, although the trend is less pronounced than that in our fiducial
run. Also, the system never achieves a quasi-steady state on its magnetic flux distribution.
We note that most previous unstratified simulations of the MRI adopt even smaller radial
domain size with $L_x=H$ (e.g., 
\citealp{HGB95,Fleming_etal00,SanoInutsuka01,LesurLongaretti07,Simon_etal09}).
Therefore, the intermittent features discussed above would make signatures of magnetic
flux concentration hardly noticeable in these simulations. We also note that the phenomenon
of magnetic flux concentration should occur in earlier unstratified simulations with
relatively large radial domain such as in \citet{Bodo_etal08} and \citet{LongarettiLesur10}.
Nonetheless, these works have mostly focused on the volume averaged turbulent
transport coefficients rather than sub-structures in the radial dimension.

Enlarging the radial domain size to $L_x=8H$, we see that the system initially develops
two ``wavelengths" of zonal structures ($t=300-1000\Omega^{-1}$), with magnetic flux
concentrated into two radial locations corresponding to the density minima. Later on,
however, the two modes merge into one single mode with much stronger density variation.
The magnetic flux in the two radial locations also merge to reside in the new density
trough. From this time, the system achieves a quasi-steady state configuration. We
find that for turbulent diffusivities, our model provides excellent fits, and the values of
$\alpha_t$, $Q'$ and $\alpha_{xy}$ agree with those in the fiducial run ID-4-16 very well.
This indicates well converged basic turbulent properties with simulation domain size, and
our phenomenological description on magnetic flux concentration works reasonably well in
a wider simulation box. On the other hand, we find that Equation (\ref{eq:alpham}) no longer
yields a good fit between the density and Maxwell stress profiles, leaving the value of
$\alpha_m$ poorly measured (the reported value represents an underestimate). This is
most likely due to the more stochastic nature of the forcing term (Maxwell stress) in a wider
simulation box, which has been discussed in \citet{Johansen_etal09}.

Further increasing the radial domain size to $L_x=16H$, we find that the system initially
breaks into multiple zonal structures. Magnetic flux still concentrates toward
low-density regions, but the density and magnetic flux profiles show long-term evolutions.
Even by running the simulation for more than 400 orbits, no quasi-steady configuration is
found. The later evolution of the system is still dominated a single ``mode" of zonal structure
in the entire radial domain, but there are more substructures associated with multiple
peaks of magnetic flux distribution. The overall level of magnetic flux concentration is
weaker, with typical $\ovl B_z^{\rm Max}/B_{z0}\sim1.5$ or less, and the typical scale of
individual magnetic flux substructure is around $\sim2H$. While we may speculate that this
simulation better represents realistic (fully-ionized) disks, we also note that the simulation
box size of this run is already large enough that the local shearing-sheet formulation would
fail if the disk is not too thin (e.g., aspect ratio $H/R\lesssim0.03$), and we have not
included vertical stratification. Overall, in the ideal MHD case, the properties of the zonal
flow and magnetic flux concentration do not converge with the box size in shearing-box
simulations.

Finally, we notice that evidence of magnetic flux concentration is already present in earlier
global unstratified simulations with net vertical flux. \citet{Hawley01} found in his simulations
the formation of a dense ring near the inner radial boundary and various low density gaps
(i.e., zonal flows) within the disk, which were tentatively attributed to a type of ``viscous"
instability.
\citet{SteinackerPapaloizou02} obtained similar results and identified the trapping of vertical
magnetic flux in the density gaps, although they did not pursue further investigation. While
shearing-box simulation results do not converge with box size, these global unstratified
simulation results lend further support to the robustness of magnetic flux concentration in
more realistic settings.

\subsection[]{Non-ideal MHD Simulations}

\begin{figure*}
    \centering
    \includegraphics[width=160mm]{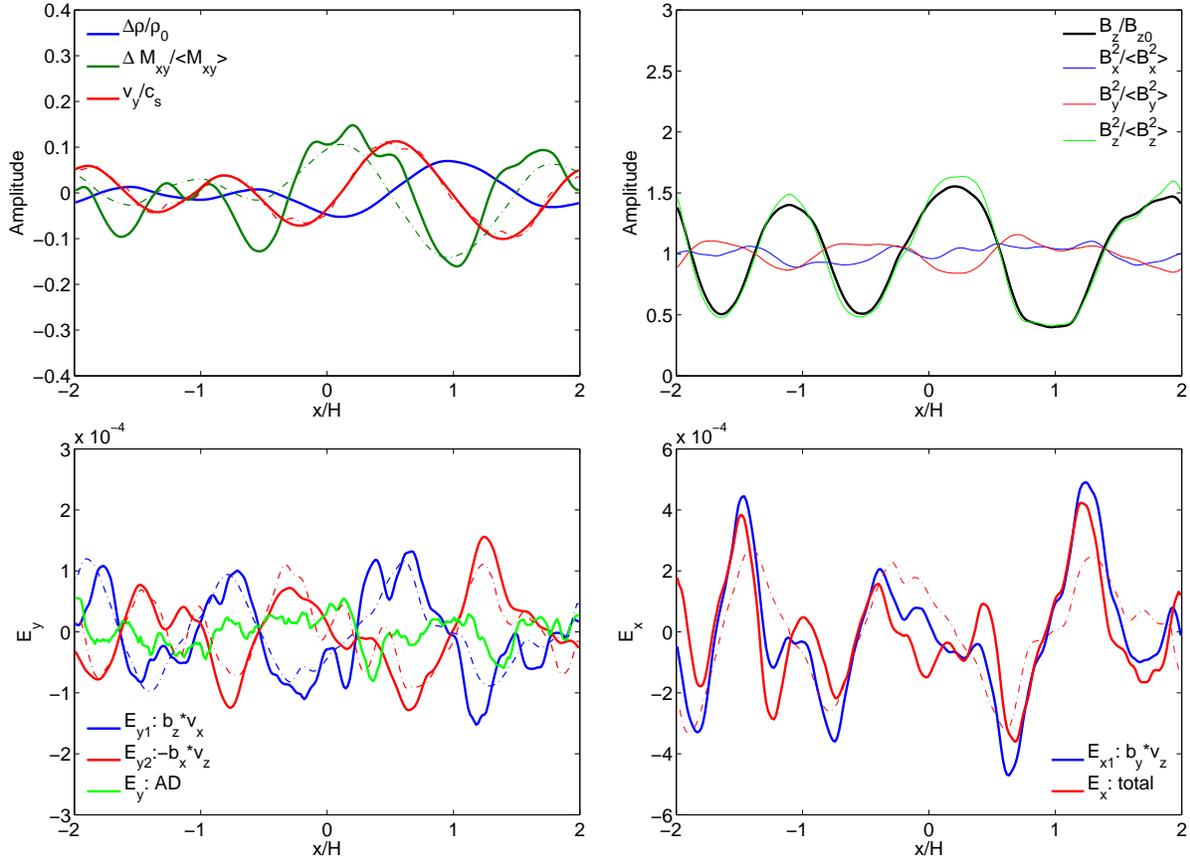}
  \caption{Radial profiles of various quantities in the saturated state of run AD-4-16, as indicated
  in the legends in each panel. Dashed-dotted lines are fits to the measured profiles based
  on the phenomenological model in Section 3. Note that the green curve in the bottom left panel
  now shows $E_y$ due to AD.}\label{fig:ADprof}
\end{figure*}

With strong AD, we first show in Figure \ref{fig:ADprof} the radial profiles of main diagnostics
from our fiducial run AD-4-16, with fitting results over plotted, which compliments our
discussions in Section \ref{ssec:niruns}. We first notice from the top right panel that because
the MRI turbulence is weaker due to AD, the mean vertical field dominates over the rms
fluctuations of the vertical field in the flux-concentrated shells. This is also the case in most
stratified shearing-box simulations for the outer regions of PPDs in \citet{Bai14}. Magnetic
fluctuations in $B_x$ and $B_y$ do not show strong trend of radial variations. 

Secondly, we find that for this run, magnetic flux concentration is still mainly due to turbulent
motions. In the bottom left panel of Figure \ref{fig:ADprof}, we also show $\ovl E_{y}^{\rm AD}$, the
toroidal electric field resulting from AD. We see that the contribution from $\ovl E_{y}^{\rm AD}$
is small compared with the other two components $\ovl E_{y1}$ and $\ovl E_{y2}$. Therefore,
the phenomenological description in Section \ref{sec:model} is equally applicable in the
non-ideal MHD case. It provides reasonable fits in the mean $\ovl E_x$ and $\ovl E_y$ profiles.
We will discuss further on the role of AD in Section \ref{sssec:AD}.

We also notice that while the radial density variation and Maxwell stress are still anti-correlated,
they are not well fitted from relation (\ref{eq:alpham}). Correspondingly, the mass diffusion
coefficient $\alpha_m$ is not very well measured. Again, this is likely due to the stochastic
nature of the forcing term (Maxwell stress). As we see from Figure \ref{fig:shbox-fid} (3rd panel
on the right), many bursts of the Maxwell stress are exerted over an extended range of the radial
domain, covering multiple peaks and troughs in the magnetic flux profile. Although on average,
regions with strong magnetic flux concentration have stronger Maxwell stress, the ``kicks" they
receive are not as coherent as its ideal MHD counterpart (3rd panel on the left). In this situation,
it is more appropriate to apply the stochastic description of the zonal flow in \citet{Johansen_etal09}
rather than the simple form of Equation (\ref{eq:massdiff}). 

\begin{figure}
    \centering
    \includegraphics[width=90mm]{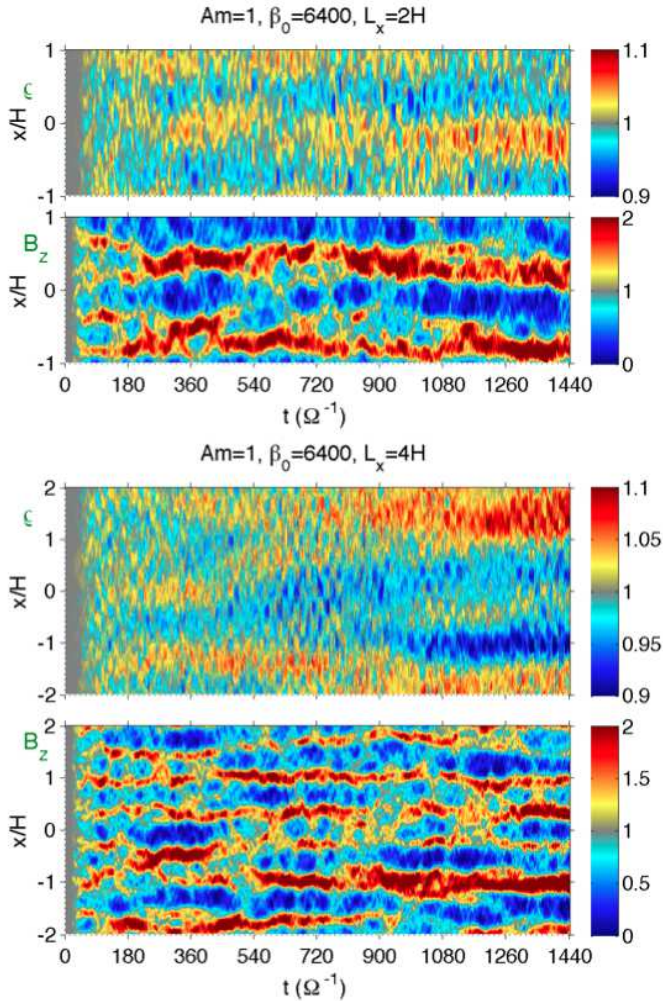}
  \caption{Time evolution of the radial profiles of $\ovl\rho$ and $\ovl B_z/B_{z0}$ from our two
  non-ideal MHD runs with different radial domain size: AD-2-64 (top) and AD-4-64
  (bottom).}\label{fig:ADhist}
\end{figure}

In Figure \ref{fig:ADhist}, we further show the time evolution of density and magnetic flux profiles
from two other runs AD-2-64 and AD-4-64 with $\beta_0=6400$ and $Am=1$.
We see that the evolutionary patterns from the two runs are very similar to each other. They are
also qualitatively similar to our run AD-4-16 discussed earlier, with magnetic flux concentrated into
thin shells whose sizes are $\lesssim0.5H$. We have also tested the results with larger box size
$L_x=6H$, and find very similar behaviors as smaller box runs. This is very different from the
ideal MHD case, and provides evidence that the properties of magnetic flux concentration converge
with simulation box size down to $L_x=2H$ in unstratified simulations. The convergence is mainly
due to the small width of the flux-concentrated shells and their small separation. Nevertheless, we
again remind the readers that properties of magnetic flux concentration and zonal flows can be
different in the more realistic stratified simulations, as mentioned in Section \ref{ssec:niruns}.

In the bottom panels of Figure \ref{fig:ADhist}, we see that the properties of the zonal flow in our
run AD-4-64 show long-term evolutions over the more than 100 orbits, and stronger density
contrast is developed toward the end of the run (which again may relate to stochastic forcing).
Similar long-term evolution behavior was also reported in unstratified simulations of \citet{Bai14}.
Despite the value of $\alpha_m$ being poorly determined, other quantities $\alpha_t$, $\alpha_{xy}$
and $Q'$ are found to be similar between the two runs AD-2-64 and AD-4-64. Their values are
a factor of $\sim2$ smaller than in run AD-4-16 with twice the net vertical field, consistent
with expectations of weaker turbulence.
In addition, we see that magnetic flux concentration is even more pronounced with weaker net
vertical field $\beta_0=6400$ than with $\beta_0=1600$. Combining the results from stratified
simulations of \citet{Bai14}, we see that strong magnetic flux concentration can be achieved with
very weak net vertical field, at least down to $\beta_0=10^5$. 

We have also computed the quality factors for these two runs with $\beta_0=6400$, and find
$\ovl Q_y\gtrsim20$ over the entire simulation domain, $\ovl Q_z\sim10-15$ in high density regions,
and $\ovl Q_z\sim20-30$ in low density regions. The small $\ovl Q_z$ value in high density regions
is mainly due to weaker (rms) vertical field, hence larger $\beta_z$, as a result of magnetic flux
concentration and zonal flows. We see that the relatively high resolution ($64$ cells per $H$ in $z$)
that we have adopted for these non-ideal MHD simulations is necessary to guarantee proper
resolution of the MRI over the entire simulation domain, especially the high-density regions of the
zonal flow.



Finally, by comparing run AD-4-16 with run ID-4-16, we see that the Maxwell stress
$\alpha_{\rm Max}$ is reduced by a factor of $\sim50$ due to AD. On the other hand, we find
that the amplitudes of $\ovl E_{y1}$ and $\ovl E_{y2}$ are reduced by just a factor of $\sim20$.
Since both $\alpha_{\rm Max}$ and $\ovl E_y$ result from quadratic combinations of
turbulent fluctuations, this fact indicates that while turbulence gets weaker, the correlation
between $v_z$ and $B_x$ becomes tighter in the AD case. To quantify this, we further define
\begin{equation}
\delta_x\equiv\frac{\langle|\ovl{v_zB_x}|\rangle}
{\langle v_z^2\rangle^{1/2}\langle B_x^2\rangle^{1/2}}\ ,\quad
\delta_y\equiv\frac{\langle|\ovl{v_zB_y}|\rangle}
{\langle v_z^2\rangle^{1/2}\langle B_y^2\rangle^{1/2}}\ ,
\end{equation}
where the overbar indicates averaging over the horizontal and vertical domain
at individual snapshots, and the angle bracket indicates further averaging over the
radial domain and selected time period in Table \ref{tab:shbox}.
For all ideal MHD runs, we consistently find that $\delta_x\sim0.07-0.09$, and
$\delta_y\sim0.09-0.11$. For all non-ideal MHD runs, we find $\delta_x\sim0.10-0.11$,
and $\delta_y\sim0.14-0.16$. It is clear that non-ideal MHD simulations give larger
$\delta$ values. In the ideal MHD case, the actual correlations between $v_z$ and $B_x$,
$B_y$ are weaker than indicated by the $\delta$ values due to stronger time
fluctuations\footnote{If we take the time average before computing the absolute
values, we obtain $\delta_{x,y}\sim0.01-0.03$ in the ideal MHD case, and
$\delta_{x,y}\sim0.04-0.07$ in the non-ideal MHD runs.}. Recently, \citet{Zhu_etal14b}
noticed that the MRI turbulence with AD has very long correlation time in vertical
velocity (see their Figures 9 and 13). The more coherent vertical motion in the AD
dominated MRI turbulence might also be related to the stronger correlation between
$v_z$ and $B_x$, $B_y$ and efficient magnetic flux concentration.

\subsubsection[]{Role of Ambipolar Diffusion on Magnetic Flux Concentration}\label{sssec:AD}

\begin{figure}
    \centering
    \includegraphics[width=90mm]{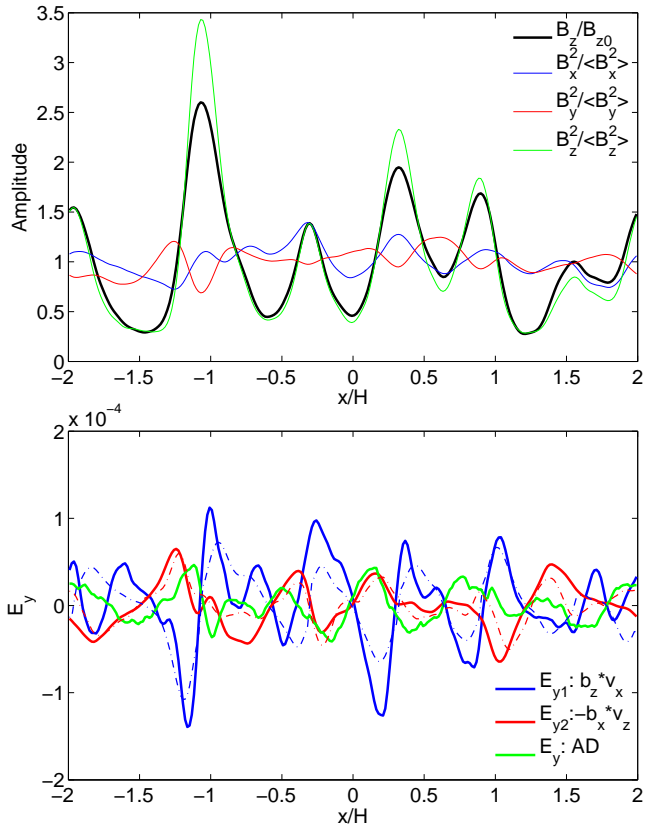}
  \caption{Equivalence of the top right and bottom left panels of Figure \ref{fig:ADprof},
  but for run AD-4-64.}\label{fig:AD64prof}
\end{figure}

As previously discussed, AD appears to play a minor role in magnetic flux concentration
in run AD-4-16, as shown in the bottom left panel of Figure \ref{fig:ADprof}. On the other
hand, we find that with weaker net vertical field as in run AD-4-64 ($\beta_0=6400$), AD
acts to enhance the level of magnetic flux concentration. In Table \ref{tab:shbox}, we see
that the value $\ovl B_z^{\rm Max}/B_{z0}$ is systematically higher in run AD-4-64
compared with run AD-4-16. In Figure \ref{fig:AD64prof}, we show the radial profiles of
$\ovl B_z$ as well as various components of $\ovl E_y$. We see that vertical flux is
squeezed into thiner shells with much sharper magnetic flux gradients compared with run
AD-4-16 (top right panel of Figure \ref{fig:ADprof}). Very interestingly, the AD electric field
$\ovl E_y^{\rm AD}$ is mostly anti-correlated with $\ovl E_{y1}$, suggesting that it plays
an anti-diffusive role, and its contribution is comparable with $\ovl E_{y2}$. We have also
checked the simulations in \citet{Bai14}, where the midplane $\beta_0=10^{4-5}$, and
found that again, contribution from $\ovl E_y^{\rm AD}$ to magnetic flux concentration is
comparable to, and sometimes more than, that from $\ovl E_{y2}$.

AD is generally thought to be a diffusive process, which tends to reduce the magnetic
field strength by smoothing out the field gradients. However, unlike Ohmic resistivity,
AD is highly anisotropic. It also preserves magnetic field topology since it represents
ion-neutral drift without breaking field lines. \citet{BrandenburgZweibel94} demonstrated
that AD can in fact lead to the formation sharp magnetic structures, especially near
magnetic nulls. This dramatic effect was attributed to two reasons. First, magnetic flux
drifts downhill along magnetic pressure gradient, and second, reduction of diffusion in
weak field regions. They also showed via a 2D example that even without magnetic
nulls, sharp current structures can be formed. While the situation is different in our
case, stronger concentration of magnetic flux with sharp vertical flux profiles can be
considered as another manifestation on the effect of AD in forming sharp magnetic
structures. Weaker net vertical field leads to weaker MRI turbulence, allowing the effect
of AD to better stand out.
%



\section[]{Discussions}\label{sec:discussion}


Our simulation results demonstrate that magnetic flux concentration and enhanced zonal
flows are robust outcome of the MRI in the presence of net vertical magnetic flux, at least
in shearing-box simulations. We have also tested the results using an adiabatic equation of
state with cooling, where we set the cooling time to satisfy $\Omega t_{\rm cool}=1$. We
find exactly the same phenomenon as the isothermal case with strong zonal flows of similar
amplitudes and strong flux concentration. Magnetic flux concentration in low density regions
of the MRI turbulence was also observed in \citet{Zhu_etal13}, where the low density region
was carved by a planet. Concentration of magnetic flux enables the planet to open deeper
gaps compared with the pure viscous case. Their results further strengthen the notion of
magnetic flux concentration as a generic outcome of the MRI turbulence.





In broader contexts, the interaction of an external magnetic field with turbulence has
been studied since the 1960s. Observations of the solar surface show that magnetic flux
is concentrated into discrete and intermittent flux tubes with intricate topology, where the
field is above equipartition strength to suppress convection. They are separated by
convective cells with very little magnetic flux. It is well understood, both theoretically and
numerically, that in a convective medium, magnetic flux is expelled from regions of
closed streamlines, and concentrates into flux tubes in between the convective cells (e.g.,
\citealp{Parker63,GallowayWeiss81,Nordlund_etal92}). In the interstellar medium, 
concentration of magnetic flux in MHD turbulence has also been suggested
\citep{Vishniac95,LazarianVishniac96}, via a process which they referred to as turbulent
pumping. Our findings are in several aspects different from the formation of flux tubes.
For example, the distribution of mean vertical magnetic field is quasi-axisymmetric rather
than patchy. Also, the level of concentration is modest, with mean vertical field typically
weaker than the turbulent field, and the overall distribution of magnetic energy is
approximately uniform. Nevertheless, our findings add to the wealth of the flux
concentration phenomena, and deserve more detailed studies in the future.


\begin{figure}
    \centering
    \includegraphics[width=90mm]{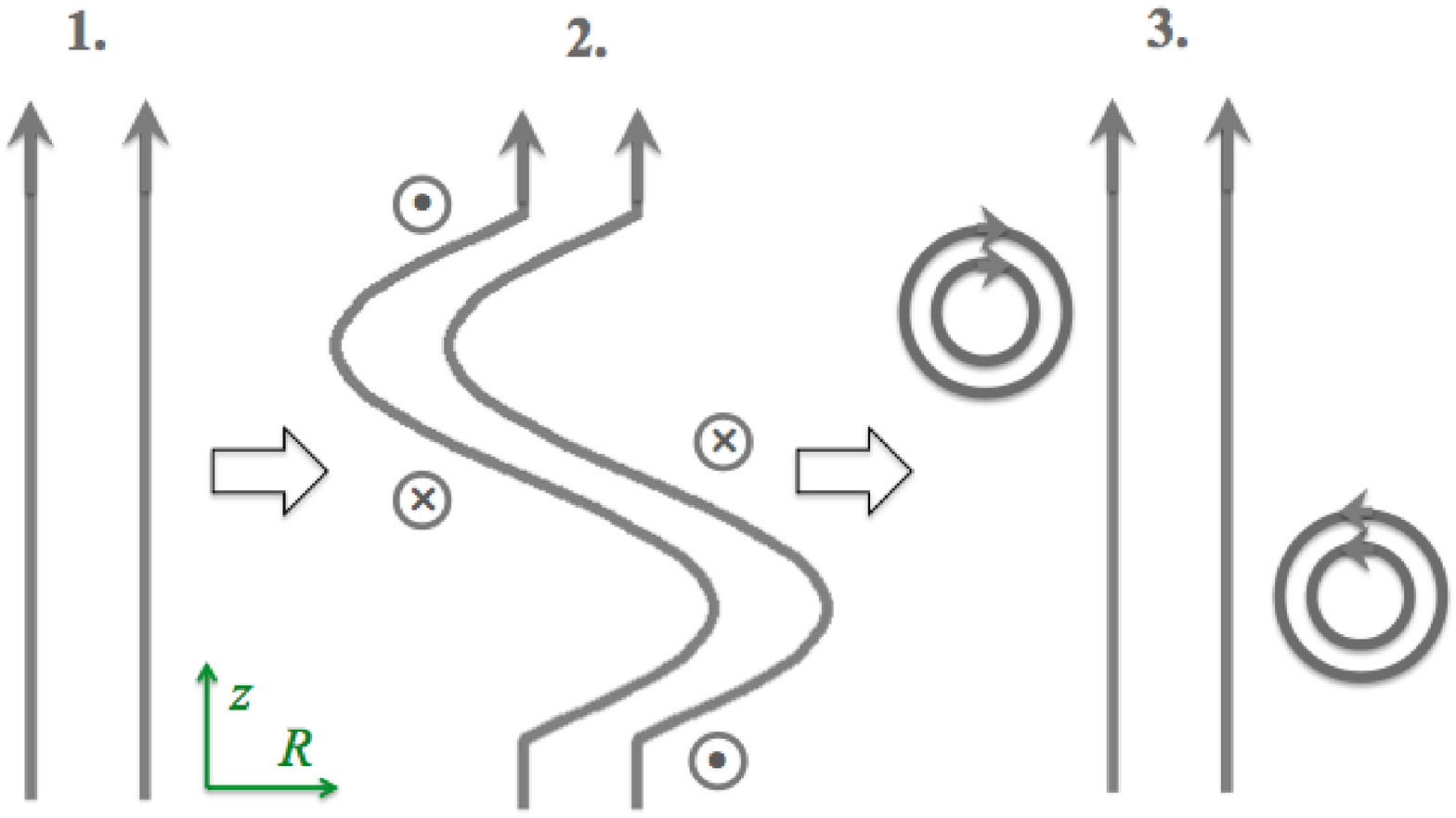}
    \includegraphics[width=90mm]{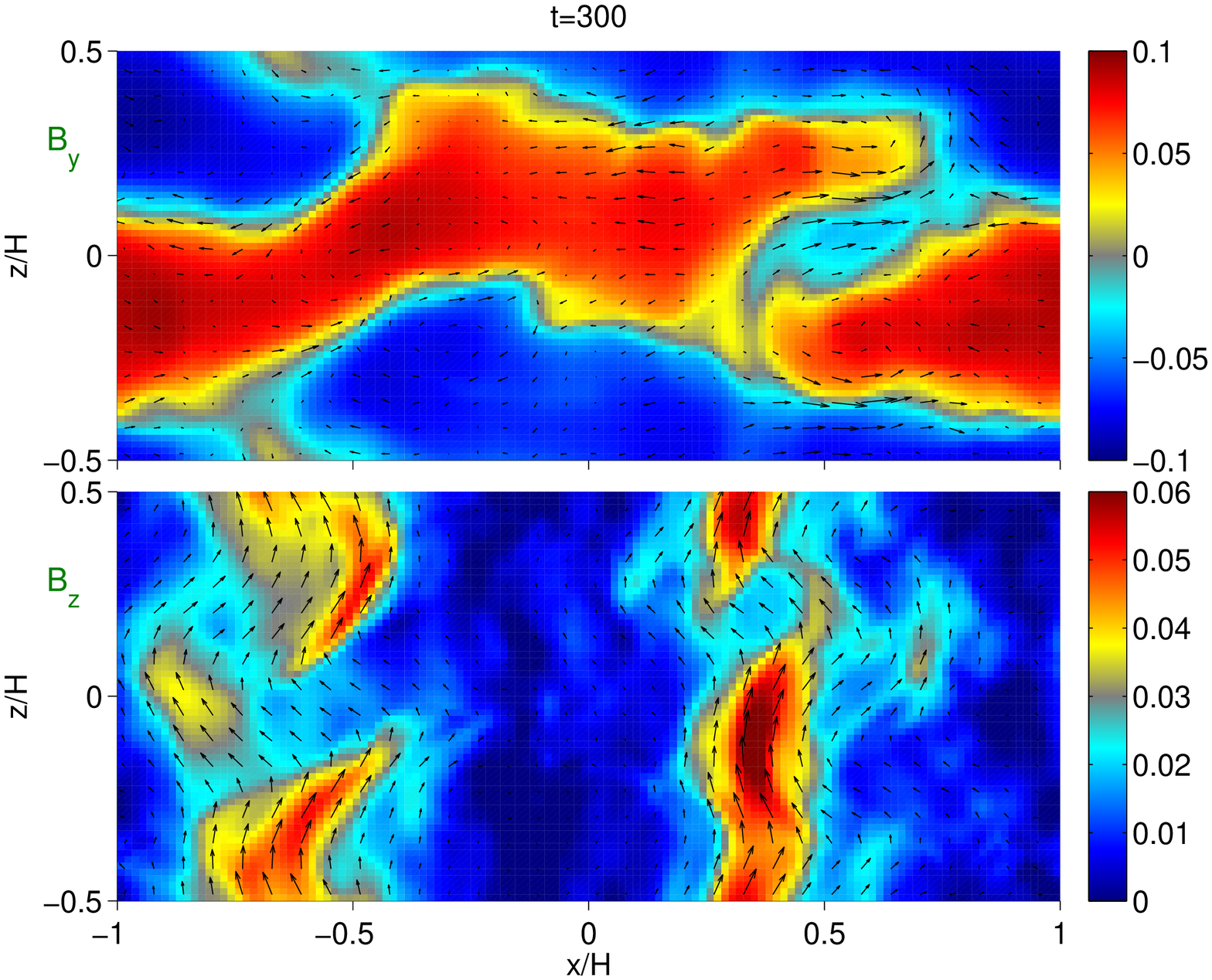}
  \caption{Top: schematic illustration of a possible mechanism for magnetic flux concentration
  due to the MRI, where the bold lines represent magnetic field lines. See explanation in Section
  \ref{ssec:physical}. Bottom: a snapshot at $t=300\Omega^{-1}$ from run AD-2-64 showing the
  azimuthally averaged toroidal (up) and vertical (down) magnetic fields. Arrows indicate the
  azimuthally averaged in-plane velocity field (up) and magnetic field (down).}\label{fig:schematic}
\end{figure}

\subsection[]{A Possible Physical Picture}\label{ssec:physical}

Here we describe a possible physical scenario for magnetic flux concentration in the MRI
turbulence. It is schematically illustrated on the top panel of Figure \ref{fig:schematic},
which is divided into three stages.

We consider the unstable axisymmetric linear MRI modes in the presence of net vertical
magnetic field (stage 1), so-called the ``channel flows". The channel flows exhibit as two
counter-moving planar streams, and are found to be exact even in the non-linear regime
\citep{GoodmanXu94}. The vertical fields are advected by the streams to opposite
radial directions, generating radial fields. The radial fields further generate toroidal fields
due to the shear. As a result, oppositely directed radial and toroidal fields are produced and
grow exponentially across each stream (stage 2). Eventually, the growth is disrupted by
parasitic instabilities or turbulence \citep{PessahGoodman09,Latter_etal09}, effectively
leading to enhanced reconnection of such strongly amplified, oppositely directed horizontal
fields around each stream. The outcome is represented by two field loops in stage 3.
Eventually, these loops are dissipated, and we are back in stage 1.

In the picture above, the material in the loop (stage 3) is originally threaded by net vertical
flux. However, due to reconnection, material is pinched off from the original vertical field
lines. Therefore, the mass-to-flux ratio in these field lines decreases.
In other words, magnetic flux is effectively concentrated into low-density regions. This
mechanism resembles the idea of turbulent pumping \citep{Vishniac95,LazarianVishniac96},
but relies on the specific properties of the MRI. In brief, the reconnection process following
the development of the channel flows effectively pumps out the gas originally threaded by
vertical field lines, which results in magnetic flux concentration.

As we have briefly discussed in Section \ref{ssec:fid1}, the evolution of the MRI shows
recurrent bursty behaviors characteristic of discrete channel flows on large scales,
followed by rapid dissipation. The overall behaviors are qualitatively similar to the cyclic
picture outlined above. A more detailed study carried out by \citet{SanoInutsuka01}
lends further support to this picture.

In the presence of strong AD, the above picture is more easily visualized since in the
flux-concentrated shells, mean vertical field dominates turbulent field. In the bottom
panels of Figure \ref{fig:schematic}, we show a snapshot of azimuthally averaged
field quantities from our run AD-2-64. We see that the flux-concentrated shells (at both
$x\sim-0.7H$ and $x\sim0.4H$) show clear signature of sinusoidal-like variations in $z$,
indicating the development of channel flows. Given the mean vertical field strength
with $\beta_0=6400$, the net vertical field in the flux-concentrated shells can be 2-4
times stronger, with $\beta_z\sim400-1600$. The corresponding most unstable
wavelength is about $0.5-1H$, consistent with observed features. In the upper panel,
we see that toroidal fields are amplified to relatively strong levels ($\beta_y\sim100$).
Oppositely directed toroidal fields are separated by sharp current sheets, ready for
reconnection to take place. Also, the location of the current sheets approximately
coincides with the location where radial field in the channel mode changes sign,
consistent with expectations.

Admittedly, the saturated state of the MRI turbulence, especially in the ideal MHD
case, contains a hierarchy of scales where the processes described above may be
taking place. The final result would be a superposition of loop formation and
reconnection at all scales. The simple picture outlined here is only meant to be
suggestive. More detailed studies are essential to better understand the physical
reality of magnetic flux concentration in the MRI turbulence.


\subsection[]{Implications for Magnetic Flux Transport}

The properties of the MRI turbulence strongly depend on the amount of net vertical
magnetic flux threading the disk \citep{HGB95,BaiStone13a}. Therefore, one key
question in understanding the physics of accretion disks is whether they possess
(or how they acquire) net vertical magnetic flux, and how magnetic flux is transported
in the disks.


Conventional studies on magnetic flux transport generally treat turbulent diffusivity
as an isotropic resistivity. Balancing viscous accretion and isotropic turbulent diffusion,
it is generally recognized that for magnetic Prandtl number of order unity (appropriate for
the MRI turbulence e.g., \citealp{GuanGammie09,LesurLongaretti09,FromangStone09}),
magnetic flux tends to diffuse outward for thin accretion disks
\citep{Lubow_etal94a,GuiletOgilvie12,Okuzumi_etal14}.

Our results indicate, at least for thin disks (where the shearing-sheet approximation is
valid), that the distribution of magnetic flux in accretion disks is likely non-uniform.
\citet{SpruitUzdensky05} showed that if magnetic flux distribution is patchy, inward
dragging of magnetic flux can be much more efficient because of reduced outward
diffusion and enhanced angular momentum loss on discrete patches. While in our study,
magnetic flux concentrates into quasi-axisymmetric shells rather than discrete bundles,
we may expect similar effects to operate as a way to help accretion disks capture and
retain magnetic flux.

Given the highly anisotropic nature of the MRI turbulence, our results also suggest that it
is important to consider the full turbulent diffusivity/conductivity tensor in the study of
magnetic flux transport. While the full behavior of this tensor is still poorly known, our
results have already highlighted its potentially dramatic effect in magnetic flux evolution.
Additionally, magnetic flux evolution can also be strongly affected by global effects, which
requires careful treatment of disk vertical structure, as well as properly incorporating
various radial gradients that are ignored in shearing-box \citep{Beckwith_etal09,GuiletOgilvie14}.

\subsection[]{Zonal Flow and Pressure Bumps}

Our results suggest that magnetic flux concentration and zonal flows are intimately connected.
In the context of global disks, radial variations of concentrated and diluted mean vertical
field lead to variations of the Maxwell stress or effective viscosity $\nu$. Steady state accretion
demands $\nu\Sigma={\rm const}$ \citep{Pringle81}. Correspondingly, the radial profile of
surface density (hence midplane gas density) is likely non-smooth. The density/pressure
variations drive zonal flows as a result of geostrophic force balance. Therefore, the enhanced
zonal flows reported in stratified shearing-box simulations with net vertical magnetic flux
\citep{SimonArmitage14,Bai14} are likely a real feature in global disks.

In PPDs, the radial pressure profile is crucial for the growth and transport of dust grains
(e.g., \citealp{Birnstiel_etal10}), the initial stage of planet formation. Planetesimal formation
via the streaming instability favors regions with small radial pressure gradient
\citep{Johansen_etal07,BaiStone10c}. Sufficiently
strong pressure variations may even reverse the background pressure gradient in localized
regions to create pressure bumps, which are expected to trap particles or even planets
(e.g., \citealp{KretkeLin12}). Numerical modelings indicate that such pressure bumps are
needed in the outer region of PPDs to prevent rapid radial drift of millimeter sized grains
\citep{Pinilla_etal12}. 

Realistic stratified global disk simulations with net vertical magnetic flux is numerically difficult
and the results can be affected by boundary conditions. Keeping the potential caveats in mind,
radial variations of surface density and Maxwell stress are present in the recent global stratified
simulations by \citet{SuzukiInutsuka14}, and pressure bumps are also observed in some of
their runs. Long-lived zonal flows as particle traps are also seen in the recent global unstratified
simulations by \citet{Zhu_etal14b}. Therefore, we speculate that because of strong magnetic flux
concentration, enhanced zonal flows have the potential to create pressure bumps in the outer
regions of PPDs.

\section[]{Conclusions}


In this work, we have systematically studied the phenomenon of magnetic flux concentration
using unstratified shearing-box simulations. In the presence of net vertical magnetic field,
the non-linear evolution of the MRI generates enhanced level of zonal flows, which
are banded quasi-axisymmetric radial density variations with geostrophic balance
between radial pressure gradient and the Coriolis force. We find that vertical magnetic
flux strongly concentrates toward the low density regions of the zonal flow, where
the mean vertical field can be enhanced by a factor of $\sim2$. High density regions
of the zonal flow has much weaker or even zero mean vertical field.

In ideal MHD, we find that strong magnetic flux concentration and zonal flow occur
when the radial domain size $L_x$ reaches $\sim4H$. The typical length scale of
magnetic flux concentration is $\sim 2H$, but the general behaviors of flux
concentration do not show clear sign of convergence with increasing simulation box
size up to $L_x=16H$. In non-ideal MHD with strong ambipolar diffusion (AD), magnetic
flux concentrates into thin shells whose width is typically less than $\sim0.5H$. AD
facilitates flux concentration by sharpening the magnetic flux profiles, especially when
net vertical flux is weak. The properties of the system converge when the radial domain
size reaches or exceeds $\sim2H$.

Concentration of magnetic flux is a consequence of anisotropic turbulent diffusivity
of the MRI. At the saturated state, a turbulent resistivity tends to smear out
concentrated magnetic flux. This is balanced by an anti-diffusion effect resulting from
a correlation of $\ovl{v_zB_x}$, which has the analogy to the microscopic Hall effect. In
addition, a correlation of $\ovl{v_zB_y}$ yields a radial electric field, mimicking the
classical Hall effect. We provide a phenomenological description that reasonably fits
the simulation results.
The physical origin of magnetic flux concentration may be related to the recurrent
development of channel flows followed by enhanced magnetic reconnection, a
process which reduces the mass-to-flux ratio in localized regions.

Systematic studies of turbulent diffusivities in the presence of net vertical magnetic
flux are crucial to better understand the onset of magnetic flux concentration,
together with its saturation amplitude. They are also important for understanding
magnetic flux transport in general accretion disks. Association of magnetic flux
concentration with zonal flows also has important consequences on the structure
and evolution of PPDs. This relates to many aspects of planet formation, especially on
the trapping of dust grains and planetesimal formation. In the future, global stratified
simulations are essential to provide a realistic picture on the distribution and
transport of magnetic flux, as well as global evolution of accretion disks.

\acknowledgments

We thank Charles Gammie, Hantao Ji, Julian Krolik, Dong Lai, Ramesh Narayan,
John Papaloizou and Zhaohuan Zhu for useful conversations, and an anonymous
referee for a useful report. X.-N.B is supported
by NASA through Hubble Fellowship grant HST-HF2-51301.001-A awarded by the
Space Telescope Science Institute, which is operated by the Association of
Universities for Research in Astronomy, Inc., for NASA, under contract NAS 5-26555.
JMS is supported by NSF grants AST-1312203 and AST-1333091.
Computation for part of this work was performed on Stampede at Texas Advanced
Computing Center through XSEDE grant TG-AST140001.


\bibliographystyle{apj}

\label{lastpage}
\end{document}